\documentclass[12pt,a4paper]{article}
\usepackage[T1]{fontenc}
\usepackage[utf8]{inputenc}
\usepackage[english]{babel}
\usepackage{amsmath,amssymb}
\usepackage{bm}
\usepackage{graphicx}
\usepackage{array}
\usepackage{geometry}
\usepackage{hyperref}
\hypersetup{colorlinks=true, linkcolor=blue, citecolor=blue, urlcolor=blue}

\newcommand{\sat}{\operatorname{sat}}
\DeclareMathOperator{\diag}{diag}
\DeclareMathOperator{\col}{col}
\DeclareMathOperator{\arctg}{arctg}

\begin{document}

\title{\large\textbf{Coordinated Trajectory Control Algorithm for Quadcopter Motion along a Smooth Spatial Trajectory}}

\author{%
    \textbf{S.~A.~Kim}$^{1}$, \textbf{A.~A.~Pyrkin}$^{1}$, \textbf{O.~I.~Borisov}$^{1}$\\[2mm]
    $^{1}$ITMO University, Saint Petersburg, Russia\\
    E-mail: skim@itmo.ru
}

\date{}

\maketitle

\begin{abstract}
A complete model of the motion of a quadcopter along a smooth spatial trajectory is presented. Based on the model, a robust algorithm is proposed for controlling a quadcopter based on measurements of linear coordinates and yaw angle. Using additional integrators, a dynamic control algorithm with a simplified controller tuning is designed.
\end{abstract}

\textbf{Acknowledgements:} This work was supported by the Ministry of Science and Higher Education of the Russian Federation, state assignment passport No.~2019-0898.\\[1mm]

\section*{Introduction}

Trajectory control can be broadly classified into two types, tracking and coordinated [1]. In the former case, the desired value of the regulated variables (position and orientation) is specified at each instant, and the control error represents the mismatch between the desired and the current values of the plant output variables [2,3,4]. This approach implies covering the shortest distance between the current and the desired positions and does not suppose the presence of moving obstacles along the path. If the shortest path passes through walls, obstacles, or forbidden zones, and unmeasurable disturbances may act on the plant, then the problem of precise spatial path planning becomes untenable. In this case, it is expedient to use the coordinated control method, where the motion trajectory and the speed of the motion along this curve are specified. Then the control error is the shortest distance from the current robot position to the desired trajectory, as well as the mismatch in linear velocity and orientation [5,6,7]. In this case, priority is given to minimal deviation from the trajectory, which involves numerous turns to avoid obstacles.

The present article addresses the problem of coordinated control of quadcopter motion along a smooth spatial trajectory using measurements of linear coordinates and yaw angle [2,3,4]. The dynamic motion model of this object in space is nonlinear, and the model of robot deviations from the desired trajectory is more sophisticated due to the emerging holonomic connection among all controllable variables [1,5,6,7]. The problem of controlling such a multivariable plant is further complicated by the fact that the relative degree is not identical in each control channel, and decomposition of the model into homogeneous subsystems is not feasible, unlike the case of dynamic positioning at a point [2,3,4].

In [2,3], the authors use an incomplete motion model that does not account for yaw angle dynamics. Moreover, the proposed tracking control algorithms assume a non-obvious tuning methodology, where the regulator parameter values must be chosen to satisfy conditions related to the proofs of statements. In [4], the identified shortcomings are addressed, resulting in a complete model of quadcopter motion for the problem of dynamic point positioning. Based on this model, a dynamic tracking control algorithm with a simplified regulator tuning methodology is synthesized, but it supposes measurement of the state vector.

The coordinated control algorithm for the motion along a smooth trajectory in terms of the geometric approach, as in [2,3,4], is presented in [7], however, it refers to the planar motion of a mobile robot. The present article addresses the problem of synthesizing a control law for quadcopter motion along a spatial curve based on measurements of the output regulated variables, using the approaches described in [4,7].

\section{Problem Statement}

Consider a Cartesian coordinate system \(OXYZ\) and a quadcopter-type robot with linear coordinates \(P = \col(x, y, z)\) and orientation (yaw, pitch, roll) \(\Theta = \col(\varphi, \theta, \psi)\), moving in space along a smooth trajectory \(S\) (Fig.~1).

The curve \(S\) can be specified as a system of equations of the form
\[
x_S = \eta_x(s),\quad y_S = \eta_y(s),\quad z_S = \eta_z(s),
\]
parameterized as the locus of points corresponding to the path coordinate \(s\). Another way to specify the curve is as an intersection of two smooth surfaces, that is, as a system of two equations
\[
\eta_1(x_S, y_S, z_S) = 0,\quad \eta_2(x_S, y_S, z_S) = 0.
\]

\textbf{Remark 1.} Noteworthily, an infinite set of pairs of functions \(\eta_1\) and \(\eta_2\) has the intersection corresponding to a spatial curve.

\begin{figure}[h]
\centering
\includegraphics[width=0.5\textwidth]{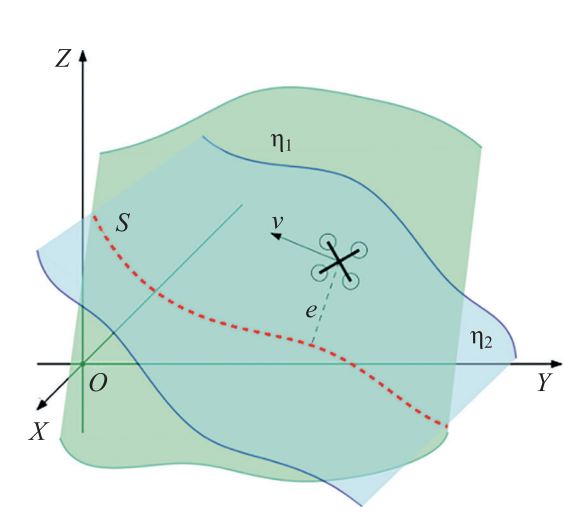}
\caption{Quadcopter motion along a spatial trajectory.}
\label{fig:1}
\end{figure}

The spatial motion model without considering friction forces has the following form [4]:
\begin{gather}
\begin{bmatrix} \dot{x} \\ \dot{y} \\ \dot{z} \end{bmatrix} = 
\begin{bmatrix} v_x \\ v_y \\ v_z \end{bmatrix}, \quad 
\begin{bmatrix} \dot{v}_x \\ \dot{v}_y \\ \dot{v}_z \end{bmatrix} = 
\begin{bmatrix} c_\phi s_\theta c_\psi + s_\phi s_\psi & 0 \\ s_\phi s_\theta c_\psi - c_\phi s_\psi & 0 \\ c_\theta c_\psi & g \end{bmatrix}
\begin{bmatrix} (u_1 + g) \\ g \end{bmatrix},\label{eq:3}\\
\begin{bmatrix} \ddot{\phi} \\ \ddot{\theta} \\ \ddot{\psi} \end{bmatrix} = 
\begin{bmatrix} u_2 \\ u_3 \\ u_4 \end{bmatrix}, \quad 
\begin{bmatrix} u_1 \\ u_2 \\ u_3 \\ u_4 \end{bmatrix} = M
\begin{bmatrix} 1 & 1 & 1 & 1 \\ 1 & -1 & 1 & -1 \\ -1 & -1 & 1 & 1 \\ -1 & 1 & 1 & -1 \end{bmatrix}
\begin{bmatrix} F_1 \\ F_2 \\ F_3 \\ F_4 \end{bmatrix} - 
\begin{bmatrix} g \\ 0 \\ 0 \\ 0 \end{bmatrix},\label{eq:4}
\end{gather}
where \((x, y, z)\) and \((\phi, \theta, \psi)\) are the linear and angular coordinates of the robot in space; \(M = \diag\bigl(\frac{1}{m}, \frac{C}{\rho}, \frac{\ell}{J_0}, \frac{\ell}{J_{\psi}}\bigr)\) are the mass-dimensional parameters; \(F = \col(F_1, F_2, F_3, F_4)\) are the forces developed by the propellers; \(U = \col(u_1, u_2, u_3, u_4)\) are auxiliary variables or virtual controls, and the notations \(c_{(\cdot)} = \cos(\cdot)\) and \(s_{(\cdot)} = \sin(\cdot)\) are introduced.

A control law \(F\) is required to ensure the robot motion along the curve \(S\) with a prescribed velocity \(V^*\), orientation \(\Theta^*(s)\), and deviation:
\begin{equation}\label{eq:5}
\lim_{t \to \infty} (\dot{s} - V^*) \leq \tilde{V}_{\max}, \quad 
\lim_{t \to \infty} (\Theta(t) - \Theta^*(s)) \leq \tilde{\Theta}_{\max}, \quad 
\operatorname{dist}(P, S) \leq e_{\max},
\end{equation}
where the conditions \(\tilde{V}_{\max} > 0\), \(\tilde{\Theta}_{\max} > 0\), \(e_{\max} > 0\) hold.

\section{Dynamic Model of Robot Deviation from the Trajectory}

The deviation of the robot from the spatial trajectory is defined using three linear and one angular coordinate:
\begin{equation}\label{eq:6}
s = \Sigma(P), \quad e_1 = E_1(P), \quad e_2 = E_2(P), \quad \delta_\phi = \phi - \phi^*(s), \quad \delta_\theta = \theta, \quad \delta_\psi = \psi,
\end{equation}
where the function \(\Sigma(P)\) determines the point on the curve \(S\) closest to the robot; the linear functions \(E_1(P)\) and \(E_2(P)\) determine the distance from the robot to two planes \(E_1(P) = 0\) and \(E_2(P) = 0\), whose intersection corresponds to the prescribed velocity vector \(V^*\) along the tangent to the curve \(S\) (Fig.~2, \textit{a}, plane \(E_1\) passes through the vector \(V^*\) parallel to the \(OZ\) axis, plane \(E_2\) passes through the vector \(V^*\) perpendicular to \(E_1\)); the prescribed yaw angle \(\phi^*(s)\) is related to the direction of the vector \(V^*_{XY}\), which is the projection of the vector \(V^*\) onto the \(OXY\) plane and defines the prescribed velocity in the horizontal plane (Fig.~2, \textit{b}). The prescribed pitch \(\theta^*(s)\) and roll \(\psi^*(s)\) are taken to be zero for all \(s\).

\textbf{Remark 2.} In the special case when the vector \(V^*\) is parallel to the \(OZ\) axis, the prescribed yaw angle \(\phi^*(s)\) must be additionally defined by the designer.

\begin{figure}[h]
\centering
\includegraphics[width=1.0\textwidth]{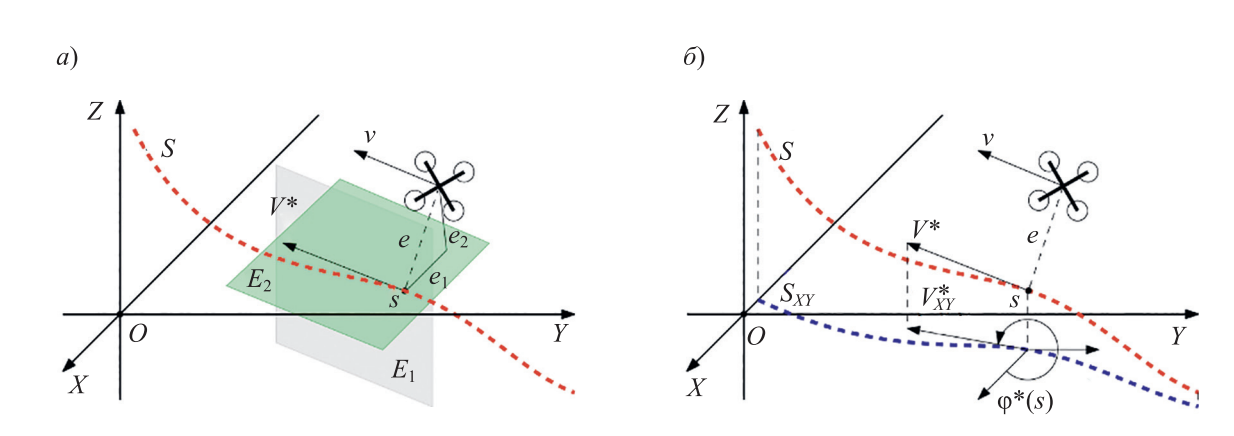}
\caption{Definition of deviations: \textit{a} — planes \(E_1\) and \(E_2\); \textit{b} — projection onto the horizontal plane.}
\label{fig:2}
\end{figure}

Let us write the kinematic relations between the regulated variables \(s\), \(e_1\), \(e_2\) and the robot coordinates \(x\), \(y\), \(z\). This requires three homogeneous transformations: one linear and two rotational:
\begin{equation}\label{eq:7}
\begin{bmatrix}
s \\
e_1 \\
e_2
\end{bmatrix}
=
R_{Y, \beta}^T R_{Z, \alpha}^T
\begin{bmatrix}
x - x_s \\
y - y_s \\
z - z_s
\end{bmatrix}
=
\begin{bmatrix}
c_\beta & 0 & s_\beta \\
0 & 1 & 0 \\
-s_\beta & 0 & c_\beta
\end{bmatrix}
\begin{bmatrix}
c_\alpha & s_\alpha & 0 \\
-s_\alpha & c_\alpha & 0 \\
0 & 0 & 1
\end{bmatrix}
\begin{bmatrix}
x - x_s \\
y - y_s \\
z - z_s
\end{bmatrix},
\end{equation}
where the first transformation corresponds to a parallel translation of the coordinate system with origin at the point \(x_s, y_s, z_s\) (the point of the curve closest to the robot), followed by a rotation about the \(Z\) axis by an angle \(\alpha = \phi^*(s)\) (the angle between the planes \(E_1\) and \(OXZ\), taking into account Remark~2) and a rotation about the \(Y\) axis by an angle \(\beta\) (the angle between the planes \(E_2\) and \(OXY\)).

Differentiating, we obtain the dynamic model for the linear coordinates:
\begin{gather}
\begin{bmatrix}
\dot{s} \\
\dot{e}_1 \\
\dot{e}_2
\end{bmatrix}
=
\begin{bmatrix}
c_\alpha c_\beta & s_\alpha c_\beta & s_\beta \\
-s_\alpha & c_\alpha & 0 \\
-c_\alpha s_\beta & -s_\alpha s_\beta & c_\beta
\end{bmatrix}
\begin{bmatrix}
v_x \\
v_y \\
v_z
\end{bmatrix},\quad
\begin{bmatrix}
\dot{v}_x \\
\dot{v}_y \\
\dot{v}_z
\end{bmatrix}
=
\begin{bmatrix}
c_\phi s_\theta c_\psi + s_\phi s_\psi \\
s_\phi s_\theta c_\psi - c_\phi s_\psi \\
c_\theta c_\psi
\end{bmatrix}
(u_1 + g) + \begin{bmatrix} 0 \\ 0 \\ -g \end{bmatrix}. \label{eq:8}
\end{gather}
For the angular deviations, we write a simplified model:
\begin{equation}\label{eq:9}
\begin{bmatrix}
\delta_\phi \\
\delta_\theta \\
\delta_\psi
\end{bmatrix}
=
\begin{bmatrix}
\phi - \varepsilon s \\
\theta \\
\psi
\end{bmatrix},\quad
\begin{bmatrix}
\dot{\phi} \\
\dot{\theta} \\
\dot{\psi}
\end{bmatrix}
=
\begin{bmatrix}
u_2 \\
u_3 \\
u_4
\end{bmatrix},
\end{equation}
where \(\varepsilon = \frac{d\alpha}{ds}\) is the curvature of the curve \(S\) projection onto the \(OXY\) plane [5].

\section{Motion Model in Normal Form}

\textbf{Statement 1.} Let
\[
u_1 = \sat_{l g}(\bar{u}_1),
\]
where \(\sat_{l g}(\cdot)\) is a smooth saturation function with level \(l g\) for any \(l \in (0, 1)\), and the variables \(\bar{u}_1\) and \(u_2\) are the outputs of two integrators with inputs \(v_1\) and \(v_2\):
\begin{equation}\label{eq:10}
\dot{\bar{u}}_1 = \rho_1, \quad \dot{\rho}_1 = v_1, \quad \dot{u}_2 = \rho_2, \quad \dot{\rho}_2 = v_2,
\end{equation}
where the variable \(u_1\) satisfies the double inequality
\begin{equation}\label{eq:11}
-l g \le u_1 \le l g.
\end{equation}
Then model (8), (9) can be represented in normal form:
\begin{gather}
\dot{\xi}_i = \xi_{i+1}, \quad i = 1, 2, 3,\\
\dot{\xi}_4 = q(\xi) + W b(\xi, \varphi) U, \label{eq:12}
\end{gather}
where \(\xi = \col(\xi_1, \xi_2, \xi_3, \xi_4) \in \mathbb{R}^{16}\) is the extended state vector; \(W\) is a nonsingular \(4 \times 4\) matrix depending on the direction and curvature of the trajectory; \(q(\xi, \varphi)\) and \(b(\xi, \varphi)\) are functions of appropriate dimensions; \(\xi_1 = \col(s, e_1, e_2, \delta_\phi)\) are the regulated variables, and \(U = \col(v_1, v_2, v_3, v_4)\) are the control inputs.

\textbf{Proof.} By sequentially differentiating the output \(\xi_1\) four times, we obtain the corresponding coordinate transformation for \(\xi_2, \xi_3, \xi_4\):
\begin{align}
\xi_2 &= \dot{\xi}_1 = W(\alpha, \beta) 
\begin{bmatrix}
v_x \\ v_y \\ v_z \\ \phi
\end{bmatrix}, \quad 
W(\alpha, \beta) =
\begin{bmatrix}
c_\alpha c_\beta & s_\alpha c_\beta & s_\beta & 0 \\
-s_\alpha & c_\alpha & 0 & 0 \\
-c_\alpha s_\beta & -s_\alpha s_\beta & c_\beta & 0 \\
-\varepsilon c_\alpha c_\beta & -\varepsilon s_\alpha c_\beta & -\varepsilon s_\beta & 1
\end{bmatrix}. \label{eq:13}
\end{align}
Note that \(\det(W)=1\) for all \(\alpha,\beta\). The inverse matrix is
\begin{equation}\label{eq:14}
W^{-1}(\alpha, \beta) =
\begin{bmatrix}
c_\alpha c_\beta & -s_\alpha & -c_\alpha s_\beta & 0 \\
s_\alpha c_\beta & c_\alpha & -s_\alpha s_\beta & 0 \\
s_\beta & 0 & c_\beta & 0 \\
\varepsilon & 0 & 0 & 1
\end{bmatrix}.
\end{equation}
Next, define \(\xi_3\):
\begin{multline}\label{eq:15}
\xi_3 = \dot{\xi}_2 = \dot{W} 
\begin{bmatrix}
v_x \\ v_y \\ v_z \\ \phi
\end{bmatrix} + W 
\begin{bmatrix}
(c_\phi s_\theta c_\psi + s_\phi s_\psi)(u_1 + g) \\
(s_\phi s_\theta c_\psi - c_\phi s_\psi)(u_1 + g) \\
c_\theta c_\psi(u_1 + g) - g \\
u_2
\end{bmatrix} \\[2mm]
= q_1(\xi) + b_1(\xi, \phi)
\begin{bmatrix}
u_1 \\
u_2 \\
\xi_3 \\
\xi_4
\end{bmatrix},
\end{multline}
where
\[
q_1(\xi) = \dot{W}\xi_2 + gT(c_\theta c_\psi - 1), \quad 
b_1(\xi, \phi) = W 
\begin{bmatrix}
0 & 0 \\
0 & 0 \\
c_\theta c_\psi & 0 \\
0 & 1
\end{bmatrix} d 
\begin{bmatrix}
c_\phi & -s_\phi \\
s_\phi & c_\phi
\end{bmatrix},
\]
\[
\dot{W} = \dot{W}W^{-1}, \quad T = W 
\begin{bmatrix}
0 \\ 0 \\ 1 \\ 0
\end{bmatrix}, \quad \xi_3 = s_\theta c_\psi, \quad \xi_4 = -s_\psi, \quad d = u_1 + g.
\]
Note that (15) is invertible because \(d \neq 0\). Direct computation yields:
\begin{align}
u_1 &= c_\theta^{-1} c_\psi^{-1} \bigl[ (s_\beta, 0, c_\beta, 0) + c_\beta (0,0,\xi_3,-s_\beta) \bigr]^T W^{-1}(\xi_3 - \overline{W}\xi_2) - g + g c_\theta^{-1} c_\psi^{-1}, \nonumber\\
\theta &= \arctg\!\left(\begin{bmatrix} 1 \\ 0 \end{bmatrix}^T \tau_3\right) = \theta(\xi_2, \xi_3, \phi), \label{eq:theta}\\
\psi &= \arctg\!\left(-\begin{bmatrix} 0 \\ 1 \end{bmatrix}^T \tau_3 \cos\!\bigl(\arctg(\begin{bmatrix} 1 \\ 0 \end{bmatrix}^T \tau_3)\bigr)\right) = \psi(\xi_2, \xi_3, \phi). \label{eq:psi}
\end{align}
At the next step, we compute \(\xi_4 = \dot{\xi}_3\) and then \(\dot{\xi}_4\), eventually obtaining (12) with
\[
q(\xi, \phi) = f_1(\xi) + f_2(\xi) \begin{bmatrix} u_1 \\ u_2 \\ \xi_3 \\ \xi_4 \end{bmatrix} + 2\dot{b}_1(\xi, \phi) \begin{bmatrix} \rho_1 \\ \rho_2 \\ \xi_3 \\ \xi_4 \end{bmatrix} + b_1(\xi, \phi) \begin{bmatrix} 0 \\ 0 \\ -s_\theta c_\psi(\dot{\theta}^2 + \dot{\psi}^2) - 2c_\theta s_\psi\dot{\theta}\dot{\psi} \\ s_\psi\dot{\psi}^2 \end{bmatrix},
\]
and \(b(\xi,\phi)\) given by
\[
b(\xi, \varphi) = 
\begin{bmatrix}
0 & 0 \\
0 & 0
\end{bmatrix} d 
\begin{bmatrix}
c_{\varphi} & s_{\varphi} \\
s_{\varphi} & -c_{\varphi}
\end{bmatrix} + \tilde{b}(\xi),
\]
with \(\tilde{b}(\xi)\) defined in the long version. Note that \(q(0,\varphi)=0\) and \(\tilde{b}(0,\varphi)=0\), and the matrix \(b(0,\varphi)\) is invertible at any time instant.

\section{Synthesis of a Robust Output-Feedback Control Law}

To synthesize a control law that ensures all three objectives (5), it is necessary to specify the desired value for the output variable. Let \(\xi_1^* = \col(V^*, 0, 0, 0)\) and \(\xi_2^* = \col(V^*, 0, 0, 0)\). Introduce the change of variables:
\begin{equation}\label{eq:19}
\tilde{\xi}_1 = \xi_1 - \xi_1^*, \quad \tilde{\xi}_2 = \xi_2 - \xi_2^*, \quad \tilde{\xi}_3 = \xi_3, \quad \tilde{\xi}_4 = \xi_4.
\end{equation}
Differentiating, we obtain
\begin{equation}\label{eq:20}
\dot{\tilde{\xi}}_1 = \tilde{\xi}_2, \quad \dot{\tilde{\xi}}_2 = \tilde{\xi}_3, \quad \dot{\tilde{\xi}}_3 = \tilde{\xi}_4, \quad \dot{\tilde{\xi}}_4 = \tilde{q}(\tilde{\xi}, \varphi, V^*) + b(\tilde{\xi}, \varphi)\overline{U},
\end{equation}
where \(\tilde{q}(\tilde{\xi}, \varphi, V^*) = q(\col(\tilde{\xi}_1+\xi_1^*, \tilde{\xi}_2+\xi_2^*, \tilde{\xi}_3, \tilde{\xi}_4)) = q(\tilde{\xi}, \varphi) + \Delta(\dot{W}, V^*)\), and the term \(\Delta\) is negligibly small for bounded curvature.

\textbf{Statement 2.} The control law of the form
\begin{gather}
\begin{bmatrix}
F_1 \\ F_2 \\ F_3 \\ F_4
\end{bmatrix}
=
\frac{1}{4}
\begin{bmatrix}
1 & 1 & -1 & -1 \\
1 & -1 & -1 & 1 \\
1 & 1 & 1 & 1 \\
1 & -1 & 1 & -1
\end{bmatrix}
M^{-1}
\begin{bmatrix}
u_1 + g \\
u_2 \\
u_3 \\
u_4
\end{bmatrix}, \quad
\begin{bmatrix}
\dot{u}_1 \\ \dot{u}_2
\end{bmatrix}
=
\begin{bmatrix}
\rho_1 \\ \rho_2
\end{bmatrix}, \quad
\begin{bmatrix}
\dot{\rho}_1 \\ \dot{\rho}_2 \\ \dot{u}_3 \\ \dot{u}_4
\end{bmatrix}
= \overline{U}; \label{eq:20a}\\
\overline{U} = b(\tilde{\xi}, \varphi)^{-1}\bigl[-\tilde{q}(\tilde{\xi}, \varphi, V^*) - \gamma_1\tilde{\xi}_1 - \gamma_2\tilde{\xi}_2 - \gamma_3\tilde{\xi}_3 - \gamma_4\tilde{\xi}_4\bigr], \label{eq:21}
\end{gather}
where the parameters \(\gamma_1, \gamma_2, \gamma_3, \gamma_4 > 0\) correspond to some Hurwitz polynomial \(\gamma(p) = p^4 + \gamma_4 p^3 + \gamma_3 p^2 + \gamma_2 p + \gamma_1\), ensures local stability of the closed-loop system and achievement of the objective (5).

\textbf{Proof.} The closed-loop model for the deviations (19) becomes
\begin{equation*}
\begin{bmatrix}
\dot{\tilde{\xi}}_1 \\
\dot{\tilde{\xi}}_2 \\
\dot{\tilde{\xi}}_3 \\
\dot{\tilde{\xi}}_4
\end{bmatrix}
=
\begin{bmatrix}
0 & I_4 & 0 & 0 \\
0 & 0 & I_4 & 0 \\
0 & 0 & 0 & I_4 \\
-\gamma_1 I_4 & -\gamma_2 I_4 & -\gamma_3 I_4 & -\gamma_4 I_4
\end{bmatrix}
\begin{bmatrix}
\tilde{\xi}_1 \\
\tilde{\xi}_2 \\
\tilde{\xi}_3 \\
\tilde{\xi}_4
\end{bmatrix},
\quad I_4 =
\begin{bmatrix}
1 & 0 & 0 & 0 \\
0 & 1 & 0 & 0 \\
0 & 0 & 1 & 0 \\
0 & 0 & 0 & 1
\end{bmatrix},
\end{equation*}
from which it is clear that if the parameters \(\gamma_i\) are chosen as coefficients of a Hurwitz polynomial, all deviation variables asymptotically tend to zero, guaranteeing (5). The local nature of stability is due to constraint (11).

However, the control law (21) requires measurements of all state variables and, within the framework of the considered problem, is not realizable. Using the extended observer method [8], the unknown terms can be eliminated. A realizable output-feedback control law, in addition to (20), also has the following form:
\begin{align}
\overline{U} &= \sat_N\!\bigl[b(0, \varphi)^{-1}\bigl(-\hat{\sigma} - \gamma_1\tilde{\xi}_1 - \gamma_2\hat{\xi}_2 - \gamma_3\hat{\xi}_3 - \gamma_4\hat{\xi}_4\bigr)\bigr]; \label{eq:U_sat}\\
\dot{\hat{\xi}}_1 &= \hat{\xi}_2 + k\alpha_1(\tilde{\xi}_1 - \hat{\xi}_1), \nonumber\\
\dot{\hat{\xi}}_2 &= \hat{\xi}_3 + k^2\alpha_2(\tilde{\xi}_1 - \hat{\xi}_1), \nonumber\\
\dot{\hat{\xi}}_3 &= \hat{\xi}_4 + k^3\alpha_3(\tilde{\xi}_1 - \hat{\xi}_1), \nonumber\\
\dot{\hat{\xi}}_4 &= \hat{\sigma} + b(0, \varphi)\overline{U} + k^4 a_4(\tilde{\xi}_1 - \hat{\xi}_1), \nonumber\\
\dot{\hat{\sigma}} &= k^5 a_5(\tilde{\xi}_1 - \hat{\xi}_1), \nonumber
\end{align}
where \(\sat_N\) is a smooth saturation function with level \(N\); the input signal \(\tilde{\xi}_1\) is defined in (19); the coefficients \(a_i\), \(i = 1, \dots, 5\), correspond to a Hurwitz polynomial \(a(p) = p^5 + a_1 p^4 + a_2 p^3 + a_3 p^2 + a_4 p + a_5\), and the gain \(k\) is greater than some number \(\kappa_0 = \kappa_0(\xi(0)) > 0\) depending on the initial conditions of the plant.

\section*{Conclusion}

This study addresses the problem of coordinated control of quadcopter motion along a smooth spatial trajectory. In contrast to [4], the problem statement of output-feedback quadcopter motion control is considered, and unlike [7] that addresses the problem of coordinated trajectory control on a plane, the present article addresses that of motion control in three-dimensional space. In the future, it is planned to conduct full-scale experimental studies of the synthesized control law.

\end{document}